# Evidence for a sub-jovian planet in the young TWA7 disk

A.-M. Lagrange[1,2]*, C. Wilkinson[1], M. Mâlin[3], A. Boccaletti[1], C. Perrot[1], L. Matrà[4], F. Combes[5], D. Rouan[1], H. Beust[2], A. Chomez[1,2], B. Charnay[1], S. Mazevet[6], O. Flasseur[7], J. Olofsson[8], A. Bayo[8], Q. Kral[1], G. Chauvin[6], P. Thebault[1], P. Rubini[9], J. Milli[2], F. Kiefer[1], A. Carter[10], K, Crotts[10], A. Radcliffe[1], J. Mazoyer[1], T. Bodrito[11], S. Stasevic[1], P. Delorme[2], M. Langlois[7]

[1] LIRA, CNRS, Observatoire de Paris, CNRS, Université PSL, 5 Place Jules Janssen, 92190 Meudon, France
[2] Univ. Grenoble Alpes, CNRS, IPAG, F-38000 Grenoble, France
[3] Steven Muller Building 3700 San Martin Drive Baltimore, MD 21218 USA
[4] School of Physics, Trinity College Dublin, the University of Dublin, College Green, Dublin 2, Ireland
[5] Observatoire de Paris, LUX, PSL University, Collège de France, Sorbonne University, CNRS; 75014 Paris
[6] Observatoire de la Côte d'Azur, Université Côte d'Azur, 96 boulevard de l'observatoire, Nice 06300, France
[7] Centre de Recherche Astrophysique de Lyon 9 Ave. Charles André 69561 Saint-Genis Laval, France
[8] European Southern Observatory, Garching bei Muenchen, Germany.
[9] Pixyl, 5 Avenue du Grand Sablon 38700 La Tronche France
[10] Space Telescope Science Institute, 3700 San Martin Dr, Baltimore, MD 21218, USA
[11] Département d'Informatique de l'École Normale Supérieure (ENS-PSL, CNRS, Inria), France

*Corresponding author(s). E-mail(s): anne-marie.lagrange@obspm.fr

## Main

Planets are thought to form from dust and gas in protoplanetary disks, and debris disks are the remnants of planet formation. Aged a few Myr up to a few Gyr, debris disks have lost their primordial gas, and their dust is produced by steady-state collisions between larger, rocky bodies[1,2]. Tens of debris disks, with sizes of tens, sometimes hundreds of au, have been resolved with high spatial resolution, high contrast imagers at optical/near-IR or (sub)-millimeter interferometers[3]. They commonly show cavities, ring-like structures, and gaps, which are often regarded as indirect signatures of the presence of planets that gravitationally interact with unseen planetesimals[2,4]. However, no planet responsible for these features has been detected yet, probably because of the limited sensitivity (typically 2-10 $M_J$) of high contrast imaging instruments (see e.g. [5,6,7,8]) prior to JWST. We have used the unprecedented sensitivity of JWST/MIRI [9,10] in the thermal IR to search for such planets in the disk of the ~6.4 Myr old star TWA 7. With its pole-on orientation, this three-ring debris disk is indeed ideally suited for such a detection. We unambiguously detected a source 1.5'' from the star, that is best interpreted as a cold, sub-Jupiter mass planet. Its estimated mass (~ 0.3 $M_J$) and position (~ 52 au, de-projected) can thoroughly account for the main disk structures.

The disk around TWA 7 is one of the youngest (6.4+/- 1 Myr, [11]) debris disks known today. TWA 7 is a close (~34 pc[12]), low-mass (0.46 $M_{Sun}$[13]) member of the young TW Hydra association, sometimes classified as a weak line, non accreting T-Tauri star [14]. The disk, resolved by HST/NICMOS[15] is one of the rare ones resolved around M stars. It is seen almost pole-on[16,17], a most favorable configuration to estimate precisely its radial distribution and to look for planets. The most recent modeling of the disk surface density deduced from VLT/SPHERE polarimetric data[17] includes a ring R1 peaking at 28 au and extending out to more than 100 au, a narrow (less than 7 au FWHM) ring at 52 au (R2), and a broader (more than 40 au FWHM) structure, R3 (93 au; Fig. 1 and 4 from [17]). No planet has been detected so far, with detection limits roughly estimated to 0.5-1 $M_J$ beyond 50 au (see supplementary information).

JWST MIRI [18] coronagraphic images in the F1140C filter (central wavelength = 11.3μm, bandwidth = 0.8μm) of TWA 7 were obtained on June, 21st, 2024 during cycle 2 (ID 3662; P.I. Lagrange). The details on the data reduction procedure are described in the Method. The main critical step is the subtraction of the residual diffracted light leaking after the MIRI coronagraph using a reference star observed with the same setup. This process is necessary to bring the contrast ratio with respect to the star to a level of $10^{-5}$ - $10^{-4}$ beyond angular separations of ~0.5''. The final image, presented in Fig. 1, reveals 3 sources within 10'' from TWA 7, the properties of which are listed in Extended Data Table 1. One source at about 4.7'' from TWA 7 (PA 107°) was classified as a stellar background source, already detected in ancillary optical HST/STIS data as well as near-IR HST/NICMOS and VLT/SPHERE data. The second one, located ~6.7'' East from TWA 7, and spatially resolved, has no counterpart (taking into account TWA 7 proper motion (-118.751 pm 0.023 mas/yr, -19.648 pm 0.026 mas/yr[19]) in the VLT/SPHERE, nor in the HST/NICMOS or STIS data. Its location in the MIRI image is consistent with that of a bright source in ALMA band 7 (346 GHz) data from 2016 [20], given the proper motion of the TWA 7 system between the ALMA and MIRI observations.This object has therefore the characteristics of a highly reddened background source. It is reminiscent of the HR 8799 multi-planetary system JWST observations for which a z~1 galaxy was detected both in the MIRI data taken at 10 and 15 micrometers, and in ALMA band 7 data as well, but never identified in the near-IR[21]. The third source, located at ~1.5'' North-West of TWA 7 (~51 au, projected separation) is unique to these MIRI observations. Hence, this source (hereafter CC#1) is extremely red, and is not compatible with any background or foreground star.

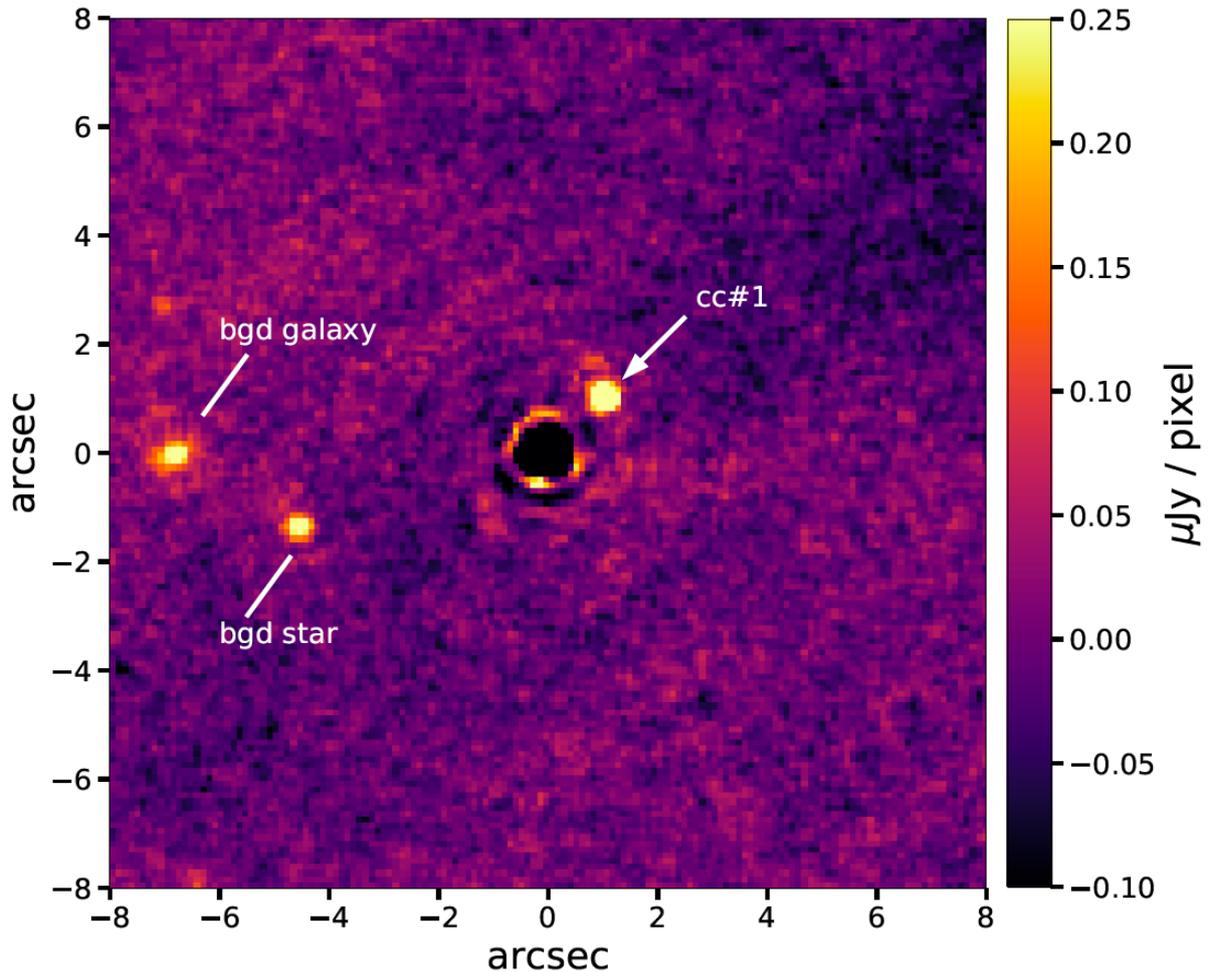

Fig. 1: JWST/MIRI image in the F1140C filter of TWA 7. North is up, and East is left. The status of three identified sources are indicated. Note that the faint signal north of the background galaxy is an artefact.

No data is available to test whether this third source shares a common proper motion with the star. In this context, we discuss in the following the possible nature of this object. The first origin one can consider is a solar system object. Yet, the vast majority of solar system objects have proper motion between 5 and 40"/hr[22]. Even very remote, low proper motion small solar system objects such as the dwarf planets Eris (sma ~68 au) or Sedna (sma ~510 au) show proper motions of 1.4''/hr[23] and 1.7''/hr[24], respectively, at the time of their discovery. No trail was observed during the 2 hour-long exposures, and no apparent motion was observed between the two images recorded during the sequences taken 2 hours apart, indicating thus that CC#1 has a proper motion less than ~0.05"/hr. A solar system object with such a low proper motion would be located at more than 200 au. For such a cold object, reflected light dominates in the MIRI F1140C filter and would require a Neptune-like size to fit the flux measured for CC#1 (for a geometric albedo of 0.1-0.3). It was checked whether such a scenario could be compatible with the hypothetical Planet Nine [25]. Based on the constraints from the planetary ephemeris [26] and the predictions of the orbit of Planet Nine [27], a solar system origin for CC#1 can be excluded definitely.

The second possible origin is a background galaxy. Like the background galaxy seen East of TWA 7, CC#1 has no reported counterpart at optical and near-IR wavelengths. But conversely to this galaxy, it has no detected counterpart in the ALMA band 7 data (see details in Supplementary Information). The detection of this unresolved source at 11.3μm, its non detection in ALMA band 7 and the measured upper limits at 1.6 μm in VLT/SPHERE data (see Supplementary Information) could still be compatible with intermediate redshift starburst galaxies. Using such galaxy templates at various redshifts and published galaxy counts in JWST fields of view, we estimated that the probability of finding one such galaxy in a 1.5" radius region centered on TWA 7 is ~0.34% (see details in Methods).

The third and last possible origin is a planet. A forward modeling approach is used to constrain the properties of this planet. Using the HADES model [28] it is possible to find fits for the JWST photometric data point while accounting for the 5-sigma upper limits provided by the high-contrast VLT/SPHERE images at 1.59 (H2) and 1.67 (H3) micrometers (Figure 2). HADES considers the thermal evolution of the planet; it assumes that the planet and the stars are coeval. Atmospheric fits incorporating water clouds indicate a narrow range, independent of other parameters, for the effective temperature between 305 and 335 K (see Extended Data Figure 3), and a mass of ~0.3$M_J$. A metallicity range above solar is required. Additional data will be necessary to further constrain this parameter.

As a comparison and extra mass estimation, evolutionary models of cold, low-mass planets [29] combined with our estimation of effective temperature are used; they lead to a comparable mass of close to 0.3 $M_J$ (see Extended Data Figure 4) for metallicities less than 2.5, as available in their framework. These two consistent results indicate that the planet mass is significantly below 1$M_J$. The current best estimate, around 0.3$M_J$, is rather insensitive to the underlying details of the two models used. It however depends on the age of the planet, assumed here to be coeval with its parent star. A younger planet would lead to a smaller mass.

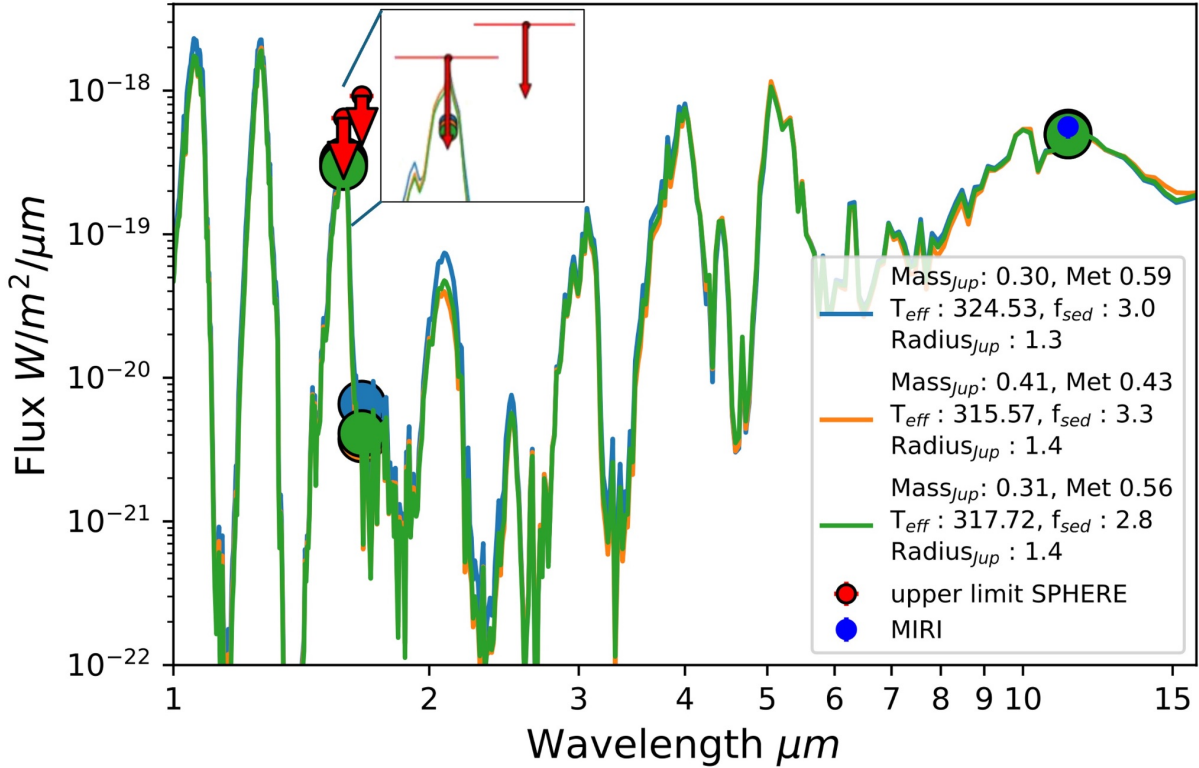

Fig. 2: Fitting of the available data on the candidate companion. 1-15μm spectra of the representative solutions fitting the observed flux of TWA 7 b in JWST filter F1140C (blue) and respecting the 5-sigma upper limits of VLT/SPHERE in filters H2 and H3 (Red), and compatible with an age of 6.4 +/- 1 Myr. The H2 and H3 respective bandwidths are 0.052 and 0.054 micrometers. A zoom around H2 and H3 is provided showing the bandwidths and the integrated points of the model spectra below the upper limit.

The observed source is located right on the R2 narrow ring, and, moreover, within a region identified as underdense compared to the rest of the ring by [17] (Figure 3a). This is very reminiscent of simulations of resonant rings predicted by early works [30,31] for closer and less massive planets, which lead [17] to envision such a possible situation for the TWA 7 system.

Numerical N-body simulations considered a 0.34 $M_J$ planet located at 52 au, a value consistent with the source measured projected separation, moving on a circular orbit and a ~13° disk inclination (assuming the planet and the disk are coplanar), and an initially flat disk of planetesimals. The planet perturbs the planetesimals, carving gaps in both sides of a narrow ring at 52 au, as well as a relative void (under density) around the planet. A top view of the resulting distribution of planetesimals after 6 Myr (shown in Figure 3b) shows a resonant planetesimal ring at 52 au, surrounded by gaps, with an under density centered on the planet location. While Figure 3a pictures the grains and Figure 3b the planetesimals, and while some differences may be expected between both distributions due to the radiation pressure acting on the grains, the simulated ring at the planet orbital radius and its surroundings are strikingly similar to the observed R2 ring and surrounding gaps. Additional information on the grains properties are needed to refine the dynamical modeling.

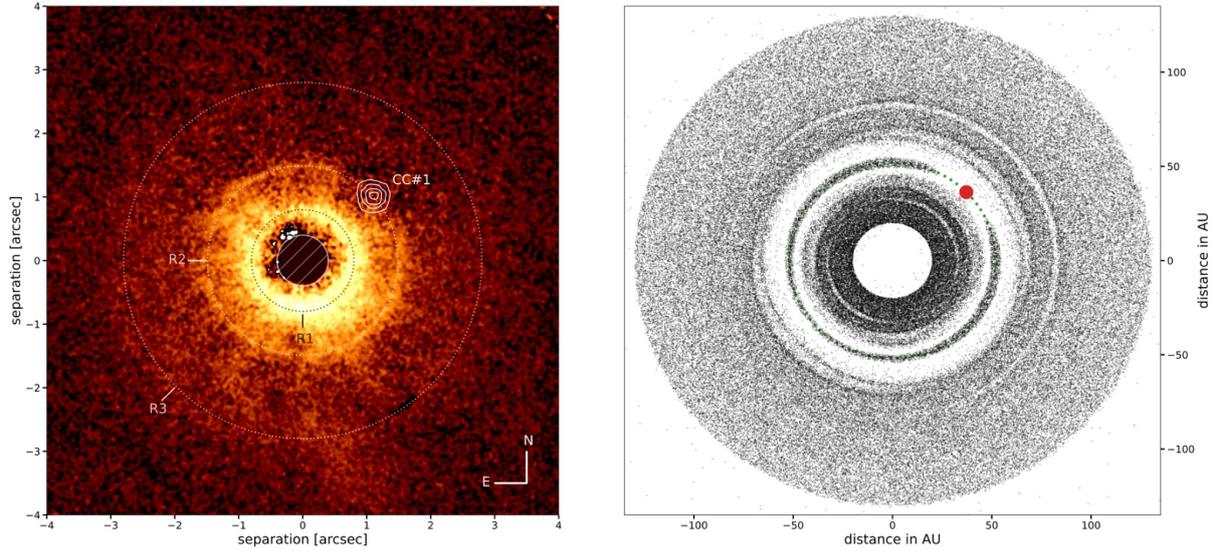

Figure 3: Image of the TWA disk and candidate companion and simulations. Left: SPHERE / IRDIS polarimetric image (in log scale) of the disk composed of the sum of 3 epochs (2016-04-26 presented in [16], 2017-03-20 presented in [17], and a new epoch 2022-02-08 reduced as in [17]), on which the MIRI image (resampled to the SPHERE pixel size) is overlaid with contours. The log of the polarimetric data is provided in Supplementary Information Data Table 2. The peak density of the rings are also indicated. The central hatched disk is a numerical mask to hide the stellar residuals. Right: Disk simulations. Top view of a disk particle with a 0.34 $M_J$ planet at 52 au, on a circular orbit, after 6 Myr. An initially flat disk of 200,000 massless particles `distributed between 20 and 130 au, with an initial surface density` $\propto r^{-1}$ was assumed. The disk evolution was computed using the symplectic N-body code swift_rmvs3[32]. The orbit of the perturbing planet is sketched in green.

The low likelihood of a background galaxy, the successful fit of the MIRI flux and SPHERE upper limits by a 0.3 $M_J$ planet spectrum, and the fact that a ~0.3 $M_J$ planet at the observed position would naturally explain the structure of the R2 ring, its underdensity at the planet's position, and the gaps, provide compelling evidence supporting a planetary origin for the observed source. Like the planet beta Pictoris b, which is responsible for an inner warp in a well resolved -from optical to mm wavelengths- debris disk[33], TWA 7b is very well suited for further detailed dynamical modeling of disk-planet interactions. To do so, deep disk images at short and millimeter wavelengths are needed to constrain the disk properties (grain sizes, etc). Refining the planet mass determination can be done with additional JWST photometry and possibly spectroscopy. Measuring the orbital parameters (eccentricity in particular) is more challenging given the long orbital period (~550 yr) of the planet. Yet, one notes that a planet on an eccentric orbit would rapidly destroy R2.

Because it is angularly well resolved from the star, TWA 7b is suited for direct spectroscopic investigations, providing thus the opportunity to study, for the first time, the interior and the atmosphere of a non-irradiated sub-Jupiter mass, cold (~320K) exoplanet, and start comparative studies with our much older and cooler solar systems giants, as well as with the recently imaged cold (~275K) but more massive (6 $M_J$) planet eps Ind Ab[34]. Improved estimations of its metallicity and temperature will further constrain its mass.

The present results show that JWST/MIRI has opened up a new window in the study of sub-Jupiter mass planets using direct imaging. Indeed, TWA 7 b (~100 $M_{Earth}$) is ten times lighter than the exoplanets directly imaged so far, and planets as light as 25-30 $M_{Earth}$ could have been detected if present at 1.5" from the star or beyond.

# Methods

1. JWST observations and data reduction and analysis

1.1 Data
Coronagraphic observations were performed with the 4QPM_1140 coronagraph paired with the F1140C filter. The details of the observations are given in Table 1. We obtained two roll angles (difference of 7.835°) to mitigate the attenuation of the coronagraph in the field of view, in case an object falls close to one phase transition of the 4QPM. Each coronagraphic observation was 2 hours long, hence a total of 4 hours on the science target. Background observations were observed immediately after the science exposures in a 2-point dithering mode, for a total of 4 hours.

A reference star, CD-23-9765, was observed back-to-back with the target in the same configuration with the aim to subtract the starlight diffraction after the coronagraph. The reference shares similar brightness and spectral type with the target, and is angularly close. It was observed with the 9-points Small Grid Dithering (SGD) to apply post-processing algorithms such as PCA [35]. In total, the reference star was observed for about 1 hour and comes with dedicated background observations.

Using comparison with simulated coronagraphic images, as in [36], we were able to estimate a pointing accuracy on the 4QPM of about 2 mas/axis, hence significantly lower than the 10 mas step of the SGD. We also confirmed the detector coordinates of the 4QPM_1440 mask (119.758, 112.158 as provided in the JWST Calibration Reference Data System).

1.2 Data analysis
The data reduction follows the steps described in [20] and in [37]. Level 1 data are retrieved from MAST (Mikulski Archive for Space Telescopes), processed with v1.14.0 of the pipeline together with CRDS files 1241. Images are registered to the coronagraph center. Calibrated files ("cal" files) are produced in house with the JWST pipeline for each roll, by subtracting the background and converting the photometric units (DN/s to MJy/sr). The background is built from the minimum per pixel of the 2 dithers. We skipped the Flat Field correction which is not appropriate for the MIRI coronagraph [21].

We took advantage of the diversity brought by the SGD mode to build a reference frame to subtract the stellar diffraction. We tested various algorithms and retained a linear combination (which uses the downhill simplex minimization) of the 9 SGD reference star images, as well as the Principal Component Analysis (PCA), as the two algorithms providing the best detection of CC#1. To mitigate the over subtraction effect, a numerical masking is implemented to ignore some parts of the image. We obtained the best compromise by selecting the annular region between 0.5'' and 3'', while the 3 sources were masked with a 1 $\lambda/D$ patch (we checked that the CC#1 flux measurements are similar with a different region: 2-3''). We proceeded similarly for the PCA, decreasing to 8 components to build the final image to be subtracted to the data in an annular region from 0.5'' to 5'' (point sources not masked). Despite the bad pixel correction applied on the raw coronagraphic images, the subtracted images with the reference star are still affected by a few bad pixels, both with the linear combination and PCA. We further apply a sigma clipping function to correct for these remaining bad pixels. Finally, the images are rotated to align the North up, considering the aperture position angles: 121.45° for roll 1 and 129.27° degrees for roll 2, as well as the V3 axis orientation on the detector (4.835°).

Next, extracting the flux and position of the CC#1 requires modeling its PSF, which can vary spatially and non-linearly due to the attenuation of the coronagraph of which the phase transitions extend across the whole field of view. We used both the diffraction code developed in [38,39] and WebbPSF [41] to simulate the MIRI PSF, taking into account the coronagraph, considering the configuration of mask, stop, and filter ('FQPM1140', 'MASKFQPM' and 'F1140C' according to the WebbPSF terminology). The position of CC#1 is approximated with a Gaussian fit, passing the sky coordinates to the former diffraction code and detector coordinates to WebbPSF in order to calculate the PSF of CC#1 accounting for the coronagraph attenuation. We measured the coronagraph transmission with both PSF estimates. The flux of the PSF model at the position of CC#1 is integrated in a 1.5$\lambda$/D aperture (to match the aperture used for photometric measurements), and compared to that at 10'' (far away from the coronagraph influence). The two approaches give similar transmissions: 0.66 and 0.62 in roll 1 and 0.31 and 0.28 in roll 2. Therefore, CC#1 is significantly closer to one 4QPM transition in roll 2 than in roll 1, so both its astrometry and photometry can be affected (Extended Data Fig. 1). We measured S/N=39 and 25, respectively, for roll 1 and roll 2. In the roll 2 image, the PSF is more asymmetrical as the planet is closer to the quadrant edge in comparison to roll 1, so we decided to consider only roll 1 data for the photometric analysis.

Based on these CC#1 PSF models, we extracted the flux and the photometry of the object by minimizing the residuals between the data and the models in a 1.5 $\lambda$/D area with 3 free parameters (positions, and flux) and using either a downhill simplex algorithm or the Nelder-Mead algorithm [40]. For comparison, we also used aperture photometry, but this required to implement an aperture correction based on simulated PSF (ratio of the total flux in the PSF to the flux integrated in the 1.5 $\lambda$/D aperture).

The flux extraction is applied both on the photometrically calibrated files (.cal) which directly provides CC#1 flux in MJy/sr, and on the uncalibrated files (.rate), which implies to measure a contrast with respect to the non-coronagraphic image of the star. As detailed in [21], the contrast measurement relies on commissioning data either on target acquisition (TA) images which come with the telescope pointing procedure but are obtained with a neutral density filter, or from images obtained on and off the coronagraph on another star. The method using TA shows some net discrepancies, likely because the TA filter is very broad (~8-18μm) while the targets have different spectral types (M3 for TWA7 and for K0/K5 for the commissioning targets). Beside, the disk emission of TWA 7 should also start to kick off beyond 15 μm. As a result, we did not use the TA method in the following. The final photometric values are based on the linear combination and PCA technique to suppress the starlight. We averaged the values of the different methods (calibrated files and contrast, aperture and PSF model for the photometric extraction) while the error bar is built from the extreme values ((max-min)/2) for being conservative. We measured a flux density of 5.60+/-0.97 e-19 W/m$^2$/μm for CC#1. The fluxes of the other sources are given in Extended Data Table 2.

2. Estimation of the probability for an intermediate-redshift galaxy

To estimate the probability that the source labeled CC#1 is a galaxy, we have taken into account the three constraints on the fluxes or upper limits at 1.6, 11 and 870 μm. The source has a flux of 22 μJy at 11 μm, and is not detected with ALMA at 870 μm, but with a tapered resolution of 2 arcsec. The measured 3-sigma upper limit for an unresolved source at the position of the MIRI source is 96 μJy[20]. Combining all ALMA observations, the 3-sigma upper limit is 76 μJy, in a beam of 0.29"x0.24" (see a detailed analysis and estimation in the Supplementary Information). The third constraint is from the non-detection at 1.6 μm with VLT/SPHERE, with a ~0.6 μJy upper limit for a point source (~60 mas in size). The limit would be much larger for an extended source, of the order of 60 μJy for a source of 0.6 arcsec. The size of a galaxy is indeed expected to vary from one wavelength to the other: at 1.6 μm the disk of old stars dominates, so the source is more extended, of the order of 10kpc; at 11μm and 870μm, the nuclear star forming region dominates, and will appear more concentrated, of the order of 100-500pc, for the same redshift.

To comply with the three constraints, if a galaxy, the source is likely to be a star-forming galaxy with a redshift between z=0.1 and 1. At lower redshifts, where 1" is smaller than 2kpc, the source would likely look extended (up to 5'') for wavelengths 1 to 4μm. At higher redshifts, the peak of the emission usually located at 100μm will enter the ALMA domain at ~1mm, and it should have been detected by ALMA. Because the star formation rate of main sequence galaxies increases considerably with redshift, star forming galaxies at redshifts between 0.1 and 1 correspond to starbursts at z=0. Therefore, we considered templates for starbursts at z=0. The starburst is in general nuclear. It can be highly peaked in the center, or distributed in a ring of ~100pc radius (like in M82), but the emitting size will be no more than typically 1 kpc. This corresponds to angular size of ~ 0.55 " at z=0.1, ~0.16 " at z=0.5, and of ~ 0.12 " at z=1.

It could also be an active nucleus, and we also considered these templates. The nucleus is then a point source at larger wavelengths, while the galaxy host is extended (~10 kpc) at the shortest wavelengths.

The probability to find a z=0.1 to 1 star forming galaxy with a flux at 11 μm equal to 22 μJy in a FOV of 10*10'' can be estimated readily through the source counts, from [42,43,44,45]. A probability of 50% is found regardless of redshift, and without any other flux constraint. To take into account the other constraints, we have used the SWIRE templates, from [46].
http://www.iasf-milano.inaf.it/~polletta/templates/swire_templates.html
which contain starburst, AGN, and normal galaxies (spiral, S0, ellipticals) at z=0, which are not representative of star-forming intermediate-redshift galaxies.
Focusing on the starbursts and AGN, we considered 14 templates. They were computed for a dozen redshifts, and calibrated all to have a flux of 22 μJy at (11μm). An illustrative plot is shown in Extended Data Figure 2, where the flux constraints are indicated by black triangles.

We used the infrared luminosity function of galaxies at redshifts between z=0.1 and z=1, in the GOODS fields, by [47], to estimate the abundance of these star-forming galaxies, as a function of their 8 μm luminosities. The latter was calculated from the templates. This lead to a probability of 12% to find a galaxy in the right redshift range in the 10*10'' FOV. The relative abundances of all these templates is not known, but we estimated that there are about 80% star forming galaxies and 20% active nuclei, including low-luminosity ones like Seyfert galaxies. As can be seen in Extended Data Figure 2, all templates comply with the constraints below z=0.5, and some starbursts are found brighter at 870 μm above this redshift. We therefore reduced the probability accordingly for z > 0.5, attributing each star forming template equal probability. The final probability for the observed source in the field of view is about 5%, with significant uncertainties +3/-2%. Hence in a radius of 1.5" around TWA7, the probability to have such a galaxy is about 0.34% (+0.22/-0.14%). We note that the distribution of galaxies on the sky might be clustered in some rare places, and our error bars should be increased somewhat, but no more than 30% given the typical errors on luminosity function of galaxies. Altogether, the probability of having a galaxy satisfying the various constraints would be 0.3% (+0.29/-0.18%).

## Extended Data

| Target name | Filter | Set up | Read-out | Groups/Int | Int./exp | Exposures/Dit | Dither | Total exposure time (s) | N roll |
|---|---|---|---|---|---|---|---|---|---|
| TWA7 (Science) | F1140C | 4QPM_1140 | FASTR1 | 500 | 60 | 1 | 1 | 7204 | 2 |
| CD-23-9765 (Reference) | F1140C | 4QPM_1140 | FASTR1 | 300 | 6 | 1 | 9 | 3894 | 1 |
| Background TWA7 | F1140C | 4QPM_1140 | FASTR1 | 500 | 60 | 1 | 2 | 7204 | 1 |
| Background Reference | F1140C | 4QPM_1140 | FASTR1 | 300 | 6 | 1 | 2 | 865 | 1 |

Extended Data Table 1. Log of observations.

| Source | Delta RA (") | Delta Dec (") | Flux W/m2/um | Resolved ? |
|---|---|---|---|---|
| CC#1 | -1.10+/-0.03 | 1.01+/0.03 | 5.60 +/- 0.97 e-19 | N |
| Star BKG | 4.49+/0.01 | -1.37+/-0.01 | 1.52 +/- 0.36 e-19 | N |
| Gal BKG | 6.74+/-0.04 | -0.05+/0.04 | 1.76 +/- 0.55 e-19 | Y |

Extended Data Table 2. Astrometry and photometry of sources detected within 10'' from TWA 7.

|  | Mass (M$_J$) | Metallicity [Fe/H] | Core mass (M$_E$) | Effective temperature (K) | Cloud sedimentation rate |
|---|---|---|---|---|---|
| Uniform priors | [0.01,2] | [-2,1.6] | [1,40] | [1,2000] | [1,5] |

Extended Data Table 3. Uniform priors used in forward modeling approach.

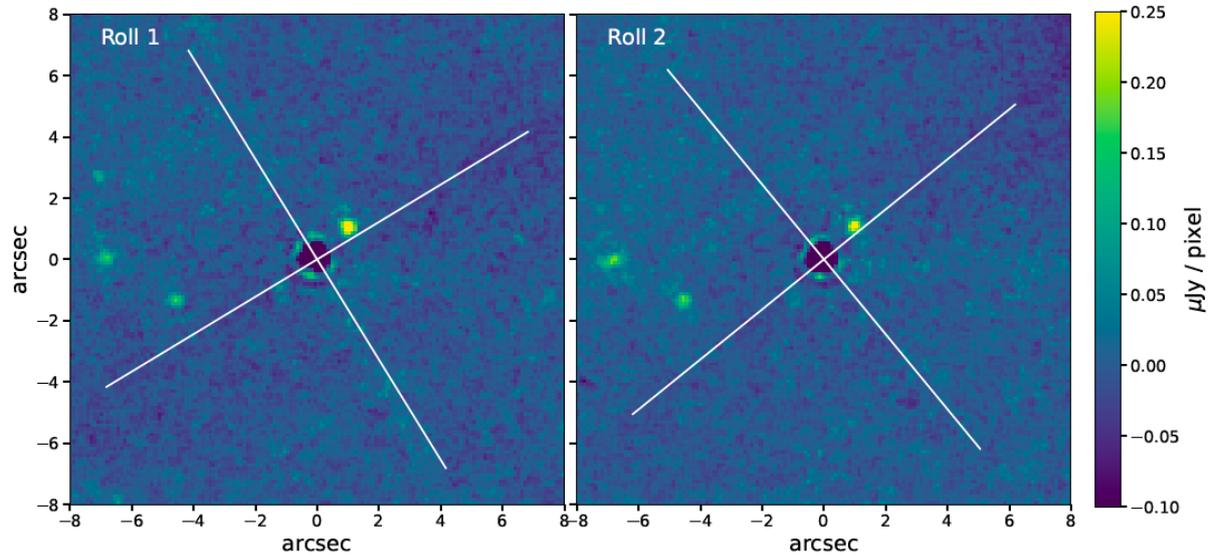

Extended Data Figure 1: JWST individual data sets. Orientation of the 4QPM and its phase transitions in the sky plane for the two telescope rolls. North ip up, East is left.

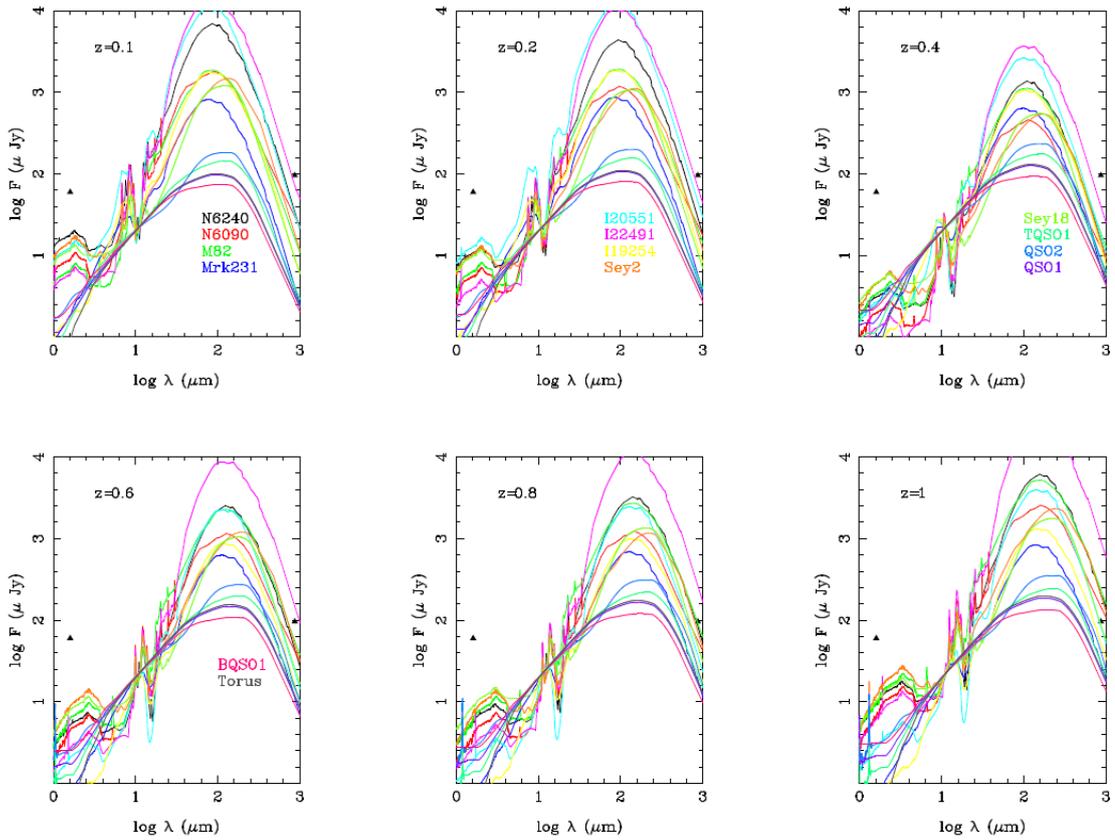

Extended Data Figure 2. SED of a set of 14 representative starburst galaxies and AGN at various redshifts. All curves are calibrated to have a flux of 22 mJy at 11 μm. The two additional constraints are marked by a black triangle: the flux should be lower than 60 mJy at 1.6 μm, and 96 mJy at 870μm.

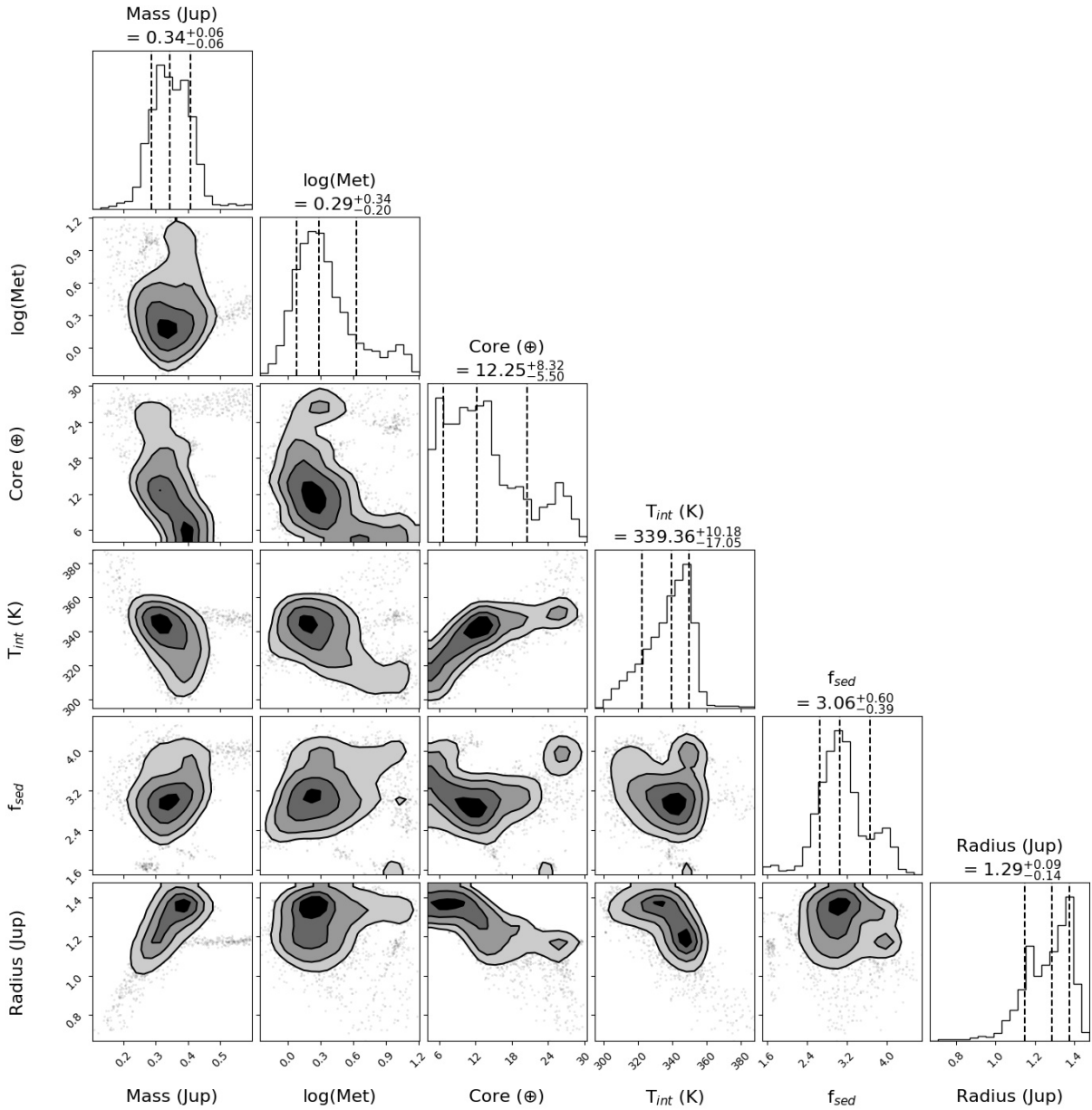

Extended Data Figure 3. Planet characterization. Corner plot of cloudy forward modeling using SPHERE upper limits and age constraints of 6.4+/- 1 Myr. The mass found is 0.34 +/- 0.06 $M_J$ (considering errors from the MCMC only). Core given in Earth mass, $T_{int}$ corresponds to the intrinsic temperature and $f_{sed}$ the sedimentation rate of the considered clouds.

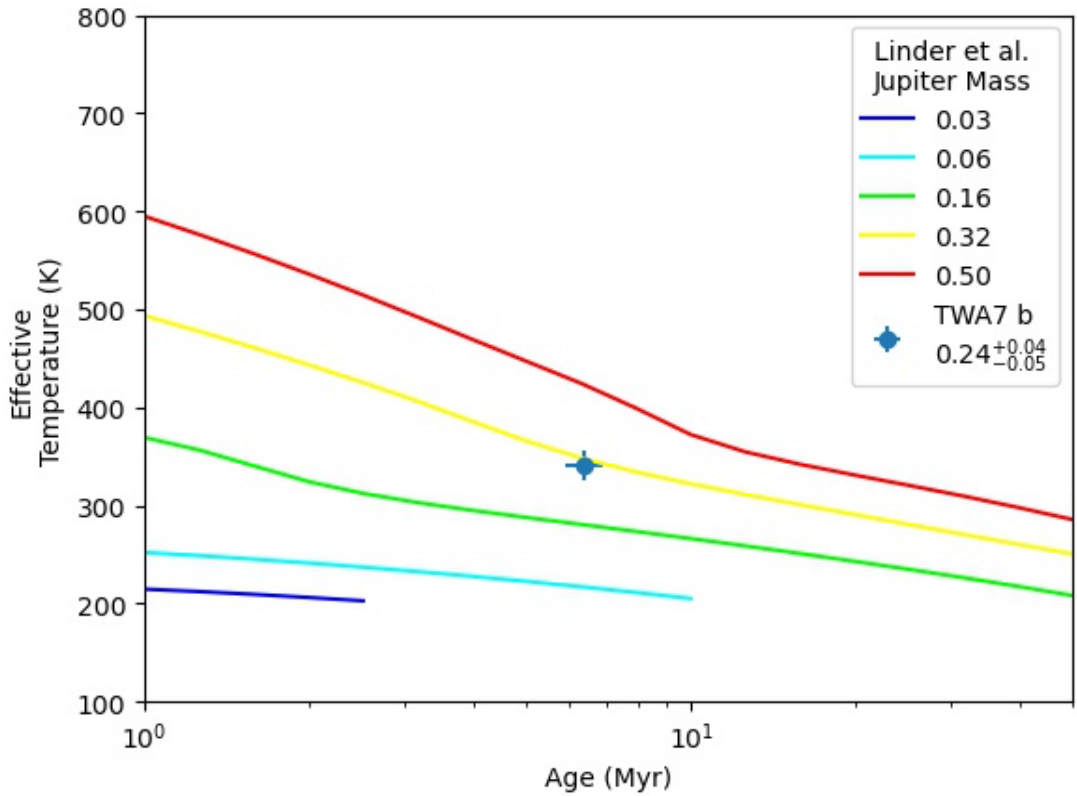

Extended Data Figure 4: Cooling models from [47]. Thermal evolution curves from [47] showing the effective temperature evolution of cloudy planets with an [Fe/H] of 0.4 dex. The estimated age and effective temperature of TWA 7b supposing the planet and star are coeval is represented by the blue point. The effective temperature used is 316+19/-23K (derived from the forward modeling) and the age 6.4+/- 1 Myr.

Research in Astronomy, Inc., under NASA contract NAS 5-03127 for JWST. These observations are associated with program #3662.

This project has received funding from the European Research Council (ERC) under the European Union's Horizon 2020 research and innovation programme (COBREX; grant agreement # 885593).

This paper makes use of ALMA data ADS/JAO.ALMA#2015.1.01015.S. ALMA is a partnership of ESO (representing its member states), NSF (USA), and NINS (Japan), together with NRC (Canada), NSC and ASIAA (Taiwan), and KASI (Republic of Korea), in cooperation with the Republic of Chile. The Joint ALMA Observatory is operated by ESO, AUI/NRAO, and NAOJ.

LM acknowledges funding by the European Union through the E-BEANS ERC project (grant number 100117693), and by the Irish research Council (IRC) under grant number IRCLA-2022-3788. Views and opinions expressed are however those of the author(s) only and do not necessarily reflect those of the European Union or the European Research Council Executive Agency. Neither the European Union nor the granting authority can be held responsible for them.

# Supplementary Informations

## 1. ALMA data analysis

We reanalysed archival ALMA data on TWA 7 from project 2015.1.01095 to search for any counterpart for CC#1 and to establish the nature of the East source detected in the MIRI images. Full details of the ALMA observations, as well as an analysis of dust and CO in the TWA 7 disk, can be found in [20,40]. In summary, the ALMA observations were taken in two antenna configurations, with one observing block in a more compact configuration on 22/04/2016, with phase center at coordinates RA: 10h42m29.904396s Dec: -33d40m17.09762s, and two observing blocks in a more extended configuration on 01/09/2016 and 15/09/2016, respectively, both with phase center at coordinates RA:10h42m30.413s Dec:-33d40m16.7s. We retrieved the pipeline-calibrated visibility datasets from the ALMA archive, and re-imaged each of the datasets with the CASA software using the tclean task, with Briggs 0.5 weighing as a compromise between resolution and sensitivity. We perform standard Hogbom deconvolution through tclean down to a residual threshold of 0.1 mJy, which corresponds to roughly 3 times the RMS noise level of each observation. Each of these single-epoch images is centered at its phase center coordinates listed above.

We repeated the same imaging process also on the combined visibility datasets, but using the tclean mosaic mode to account for the different phase centers of the different observations, so that effectively the datasets are combined on a shared coordinate system. As the phase center coordinates were offset by ~6" mostly along the RA direction between the compact and extended configuration, this creates an image with an elongated primary beam response of the ALMA interferometer. This combined image is shown in Supplementary Information Data Fig. 1, which also shows the expected proper-motion corrected location of TWA 7 (from Gaia DR3,2023j) on each of the three ALMA 2016 epochs, as well as in the MIRI epoch. Finally, Supplementary Information Data Fig. 1 shows the expected location of source CC#1, which is undetected in the ALMA data, if it had not moved compared to the MIRI location over the 8 years between observations. Similarly, we show the expected location of the MIRI East source, again assuming that it has not moved between 2016 and 2024. We find qualitatively good agreement between the MIRI East source location and the bright background galaxy detected in this combined image. The combined image has an RMS of 23 μJy/beam for a beam size (resolution element) of 0.19"x0.18". To maximize point source sensitivity at the expense of angular resolution, we also use the same method to create a separate image with natural weighting, which has an RMS of 21 μJy/beam for a beam size of 0.29"x0.24". In this natural weighting image, the 3-sigma upper limit on the flux density of a point source at the same location as found for CC#1 in the MIRI image, when corrected for the primary beam response at that location, is 76 μJy.

Note that, as explained in [20], this image does not show the disk because it is too extended and faint; it is then interferometrically 'filtered out' by these ALMA configuration observations. This filtering only affects extended emission and will not affect any signal from compact sources at the location of the disk. Significant up weighing of the short baselines of the visibility data, causing a significant degradation of the angular resolution, is necessary to bring out the faint disk emission[20]. The SNR on the disk is consequently too low to produce sufficiently accurate constraints on the location of the TWA star for comparison with the MIRI data. Nonetheless, the disk's emission centroid was found to be consistent with being centered on the expected 2016 location of TWA 7, both for dust continuum and CO[20,40].

We then focused on the bright ALMA source. To formally evaluate its consistency in location with the MIRI East source, we first measure its RA and Dec in each of the 3 epochs of observations by fitting a point source model with a flux density, RA offset and Dec offset from phase center to the visibility data. We did this using MCMC through the emcee package, with uniform, uninformative priors on each of the fitted parameters. For the uncertainties, we quadratically added the statistical uncertainty from the MCMC fit with the expected level of the ALMA systematic error for astrometric measurements. The latter comes from inaccuracies in phase referencing between the target and the phase calibrator, and is estimated to lead to systematic offsets of order 5% the beam size of each observation (https://help.almascience.org/kb/articles/what-is-the-absolute-astrometric-accuracy-of-alma). The resulting RA and Dec (Supplementary Information Table 1) of the East source is consistent across the three ALMA epochs; we derive the combined ALMA position and its statistical uncertainty from the joint posterior probability distribution of the MCMC fits to the three epochs. We then once again quadratically add the expected ALMA astrometric systematic (5% of the beam of the combined observation, i.e. 9 mas) to the statistical uncertainty, leading to the final combined ALMA coordinate for the East source.

The visibility fits yield best fit point source flux densities that are very different between the compact and extended configuration observations. As shown by imaging the residual visibilities, this is because the source is extended, with extended emission clearly present in each residual image after point source subtraction. To investigate this further, we fit a 2D Gaussian to the combined image, finding best fit FWHMs of 0.35"x0.26" with a PA of 101 degrees, and a spatially integrated flux of 4.0+/-0.4 mJy. While this model is better at capturing the resolved emission, significant residuals are still present when subtracting this best fit from the image. This suggests a more complex radial surface brightness distribution for this galaxy, with a strongly peaked core and a faint halo of emission. Regardless of the precise morphology of this galaxy, we use our best-fit RA and Dec from point source fits for the MIRI comparison as they remain an accurate estimate of the photocenter, or surface brightness centroid, of the galaxy.

Assuming the source to be extragalactic, we expect its position not to change between the ALMA and MIRI epochs. We can therefore translate the ALMA galaxy position (fixed in time, from Supplementary Information Table 1) into an offset from the (Gaia DR3-predicted) position of the TWA 7 star at the MIRI epoch, to obtain the expected offset of this galaxy in the MIRI image. This results in stellocentric RA and Dec offsets at the MIRI epoch listed at the bottom of Supplementary Information Table 1, which can be directly compared with the MIRI East source (Gal BKG) offsets in Extended Data Table 2. We find the offsets to be entirely consistent with one another within the uncertainties, and therefore firmly associate the MIRI East source to the background galaxy detected by ALMA.

2. VLT/SPHERE upper limits

The VLT/SPHERE images of TWA 7 used in the present paper to set upper limits on the candidate companion flux were obtained in the course of the SpHere INfrared survey for Exoplanets (SHINE[48]), a survey of more than 400 young, nearby stars using the guaranteed observing time (GTO) allocated to the Spectro Polarimeter High-contrast Exoplanet REsearch (SPHERE[49]) consortium. TWA 7 was observed on February 2, 2017 using the standard high-contrast coronographic sequence in pupil-stabilized mode with simultaneous observations by the infrared dual-beam imager and spectrograph (IRDIS[50]) and the infrared integral field spectrograph (IFS[51]). The H2-H3 filter pairs ($\lambda_{H2}$ = 1.593 ± 0.055 μm; $\lambda_{H3}$ = 1.667 ± 0.056 μm) were used for IRDIS and the YJ (0.95 – 1.35 μm) bands for IFS with a spectral resolution of $R_\lambda$ = 54. With a separation of 1.5", TWA 7b falls outside the IFS field of view. A first analysis of TWA 7 at H2 and H3 using classical algorithms can be found in [48]. As a next step, a more advanced algorithm, PAtch COvariance (PACO [52,53,54]) was used to reanalyse the SHINE data. PACO underlying models capture the noise statistics at the local scale of small patches through mixtures of scaled multivariate Gaussians. It provides a better estimate of the spatial and spectral correlations of the noise than classical methods. As a result, the contrasts in the IRDIS H2-H3 data were improved by more than one to two magnitudes, depending on the angular separation, compared to classical algorithms. The 5-sigma upper limits for at the position of TWA 7b, 6.4e-19 W/m2/um and 9.2e-19 W/m2/um for respectively H2 and H3 used in the present paper, were directly deduced from the contrasts computed using the approach described in [54]. These contrast curves can be found in Supplementary Information Data Fig. 2.

| Epoch | RA (hhmmss) | Dec (ddmmss) |
|---|---|---|
| 22/04/2016 | 10h42m30.4117s ± 0.0015s | -33d40m16.701s ± 0.019s |
| 01/09/2016 | 10h42m30.4130s ± 0.0006s | -33d40m16.711s ± 0.007s |
| 15/09/2016 | 10h42m30.4135s ± 0.0005s | -33d40m16.710s ± 0.005s |
| Combined | 10h42m30.4130s ± 0.0012s | -33d40m16.709s ± 0.011s |
|  | Delta RA*cos(Dec) (") | Delta Dec (") |
| Galaxy offsets from star at MIRI epoch | + 6.790 ± 0.015 | + 0.002" ± 0.011" |

Supplementary Information Data Table 1. Best-fit RA and Dec coordinates of the East (Gal BKD) source in each of the ALMA 2016 epochs, and from the combined epochs (top three rows). These were derived from point-source fits to the ALMA visibilities for each epoch; specifically they are reported as the 50+34-34th percentiles of the posterior probability distributions of the parameters, with the ALMA systematic added in quadrature. The final row reports the stellocentric offset of the East (Gal BKD) source from the stellar position at the MIRI epoch, and is therefore directly comparable with the last row of Extended Data Table 2.

| Night | Filter | Program ID | Eff. exposure time (min) | Seeing (arcsec) | Coherence time (ms) |
|---|---|---|---|---|---|
| 2016-04-26 | BB_J | 097.C-0319(A) | 49 | 0.9 | 2.9 |
| 2017-03-20 | BB_H | 198.C-0209(F) | 100 | 0.8 | 5.3 |

| 2022-02-08 | BB_H | 105.209E.001 | 119 | 0.7 | 7.3 |

Supplementary Information Data Table 2. SPHERE / IRDIS datasets used to produce the polarized intensity image of the disk shown in Fig. 3. The individual epochs were retrieved from the ESO archive, reduced following the methodology described in [17], and the three resulting images were summed up.

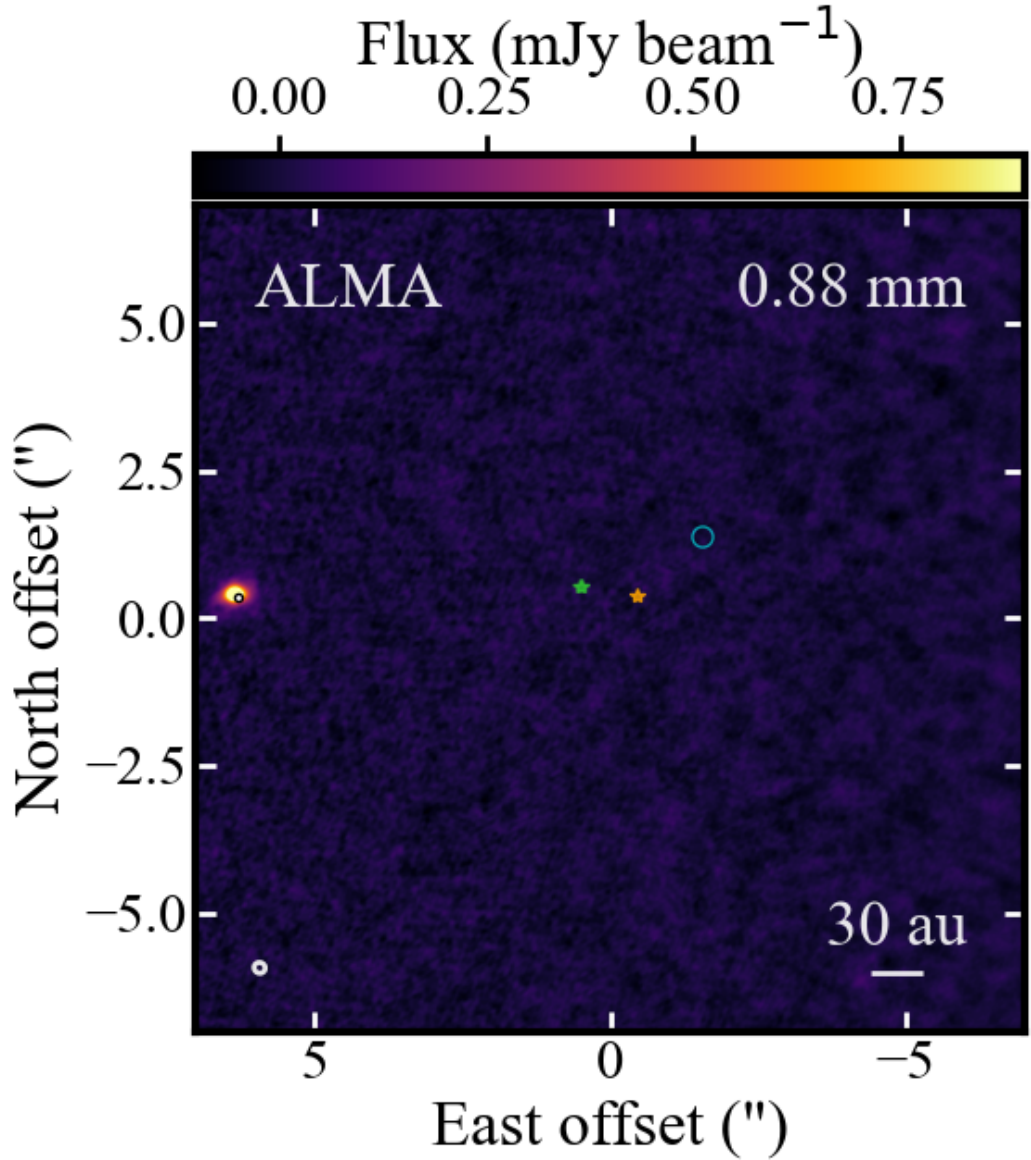

Supplementary Information Data Fig. 1. ALMA image of TWA7. Combined ALMA 0.88 mm image of the TWA 7 system obtained with Briggs 0.5 weighting, centered at the phase center of the April compact configuration observations. The background galaxy is clearly detected East of the expected stellar position, at a position consistent with the East source detected by MIRI (black and white circle). The stellar location at each of the 2016 ALMA epochs is shown by the green star (positions largely overlapping), whereas the stellar location at the 2024 MIRI observation is shown by the orange star. The position of the CC#1 source at the epoch of the 2024 MIRI observation is shown by the cyan circle. The image (not primary beam-corrected) has a resolution of 0.19" x 0.18" (shown as the circle in the bottom left of the image) and an RMS noise level of 23 µJy/beam.

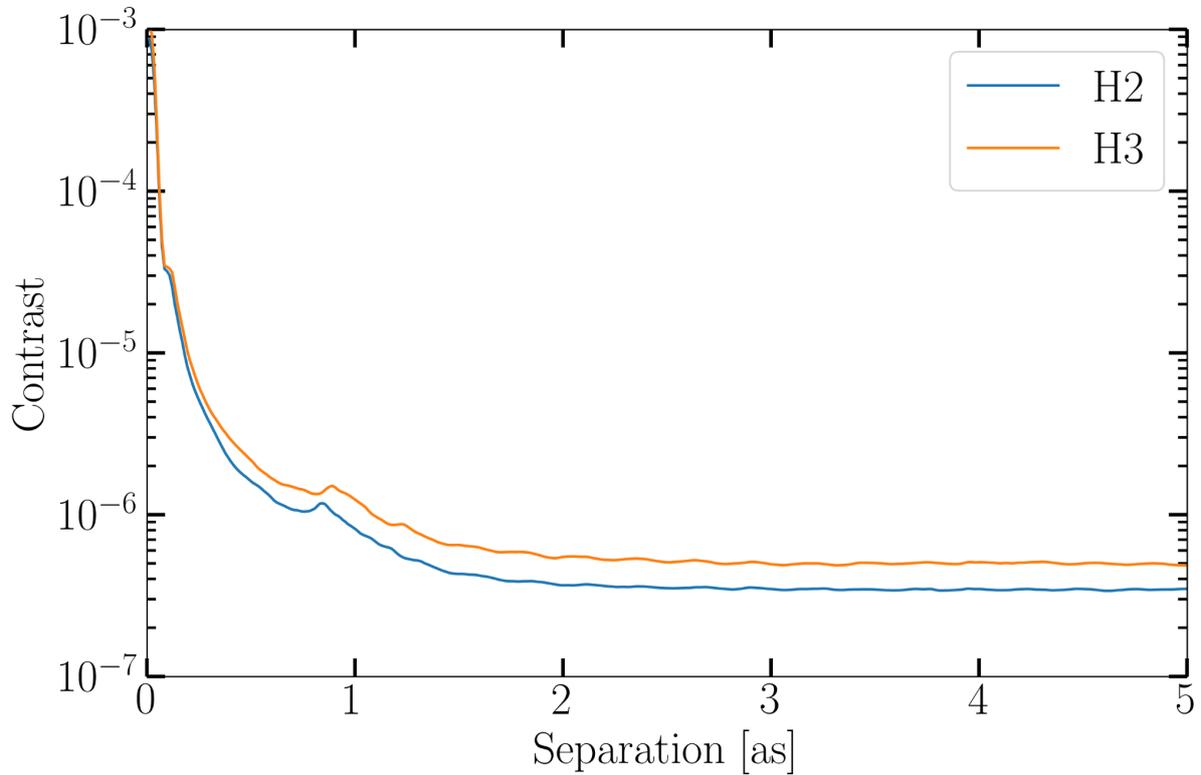

Supplementary Information Data Fig. 2. SPHERE detection limits. Contrast 5-sigma confidence level curves of the SPHERE observation used to compute the upper limits on the candidate companion flux.